\documentstyle[seceq,twocolumn,epsbox]{jpsj}

\def\runtitle{Dynamical Susceptibility in KH$_2$PO$_4$-type Crystals
above and below $T_{\mathrm c}$}
\def\runauthor{Koh {\sc Wada}, Shun-ichi {\sc Yoshida} and Norihiro {\sc
Ihara}} 

\title
{
Dynamical Susceptibility in KH$_2$PO$_4$-type Crystals above and below
$T_{\mathrm c}$ 
}

\author
{
Koh {\sc Wada}\footnote{E-mail: kwada@statphys.sci.hokudai.ac.jp}, 
Shun-ichi {\sc Yoshida}\footnote{E-mail:
shun1@statphys.sci.hokudai.ac.jp} and  
Norihiro {\sc Ihara}
}

\inst
{
Division of Physics, School of Science, Hokkaido University, Sapporo
060-0810.
}

\recdate
{
\today
}

\abst
{
The time dependent cluster approximation called the path probability
method (PPM) is applied to a pseudo-spin Ising Hamiltonian of the
Slater-Takagi model for KH$_2$PO$_4$-type hydrogen-bonded ferroelectrics
in order to calculate the homogeneous dynamical susceptibility
$\chi(\omega)$  above and below the ferroelectric transition temperature
$T_{\mathrm c}$. Above the transition temperature all the calculations
are carried out analytically in the cactus approximation of the
PPM. Below the transition temperature the dynamical susceptibility is
also calculated accurately since the analytical solution of spontaneous
polarization in the ferroelectric phase can be utilized. When
the temperature is approached from both sides of the transition
temperature, only one of relaxation times shows a critical slowing down
and makes a main contribution to the dynamical susceptibility. The
discrepancy from Slater model (ice-rule limit) is discussed in
comparison with some experimental data.
}

\kword
{
KDP(KH$_2$PO$_4$), phase transition, CVM (cluster variation method),
PPM (path probability method), dynamical susceptibility
}

\begin{document}
\sloppy
\maketitle

\section{Introduction}
Recently, we successfully applied the cluster variation method
(CVM)~\cite{KikuchiC} to the Slater-Takagi model~\cite{Takagi} for
KH$_2$PO$_4$(KDP)-type hydrogen-bonded ferroelectrics above and below
the transition temperature~\cite{Wada} to explain the anisotropy of the
wave-number dependent susceptibility $\chi(\mib{q})$ observed in the
neutron scattering experiment~\cite{Havlin}. On the other side, the path
probability method (PPM)~\cite{KikuchiP} devised by Kikuchi is the time
dependent cluster variation method and has been applied to various phase
transitions and transport phenomena~\cite{Wada2}.  Its characteristic is
that the stationary solution of the kinetic equation given by the  PPM
yields the equilibrium solution obtained from the CVM in the
corresponding approximation. Further, since the PPM provides a
systematic approximation for the kinetic problem, it makes possible to
calculate the dynamical susceptibility beyond the usual molecular field
approximation.

A few years ago Matsuo {\it et al.}~\cite{Matsuo} re-examined  the
excess entropy obtained from their own data and the other experimental
data of heat capacity for KDP. They disccused about the discrepancy from
the ice-rule of Slater model~\cite{Slater} and emphasized the significance of
excitation level in the Slater-Takagi model.

The present purpose is to calculate the dynamical susceptibility for KDP
based on the Slater-Takagi model and to compare our  results with
experimental data of the dynamical susceptibility over all the
temperature regime. Though Yoshimitu and Matubara~\cite{Yoshimitu} have
already calculated the dynamical susceptibility for the essentially same
model for the KDP above the transition temperature, their calculation
seems to be limited to the paraelectric phase. In the present paper, not
only in the paraelectric phase  but also in the ferro-electric phase we
calculate accurately the dynamical susceptibility for KDP  based on the
above mentioned Slater-Takagi model by making use of an analytical
expression for the spontaneous polarization~\cite{Ishibashi}.

\section{Formulation}
\begin{figure}[tbp]
\vspace*{0.5cm}
\begin{center}
 \epsfile{file=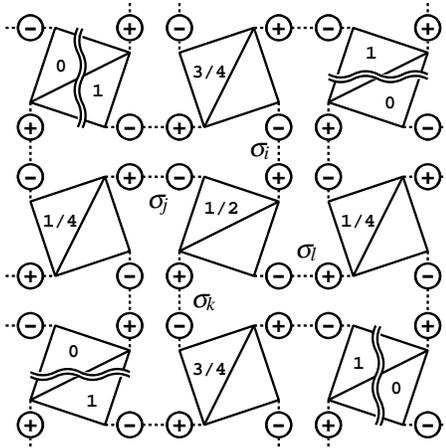,width=6cm}
\end{center}
\caption{$z$-axis projection of hydrogen bonds connecting PO$_4$
 complexes and $\sigma_i,\,\sigma_j,\,\sigma_k,\,\sigma_l$ showing the
four different pseudo-spins for protons around a PO$_4$ tetrahedron.
The numeral in the center of each PO$_4$ tetrahedron shows relative 
heights of PO$_4$ along $z$-axis.}
\label{fig:KH2PO4}
\end{figure}

\begin{figure}[tbp]
\vspace*{0.5cm}
\begin{center}
 \epsfile{file=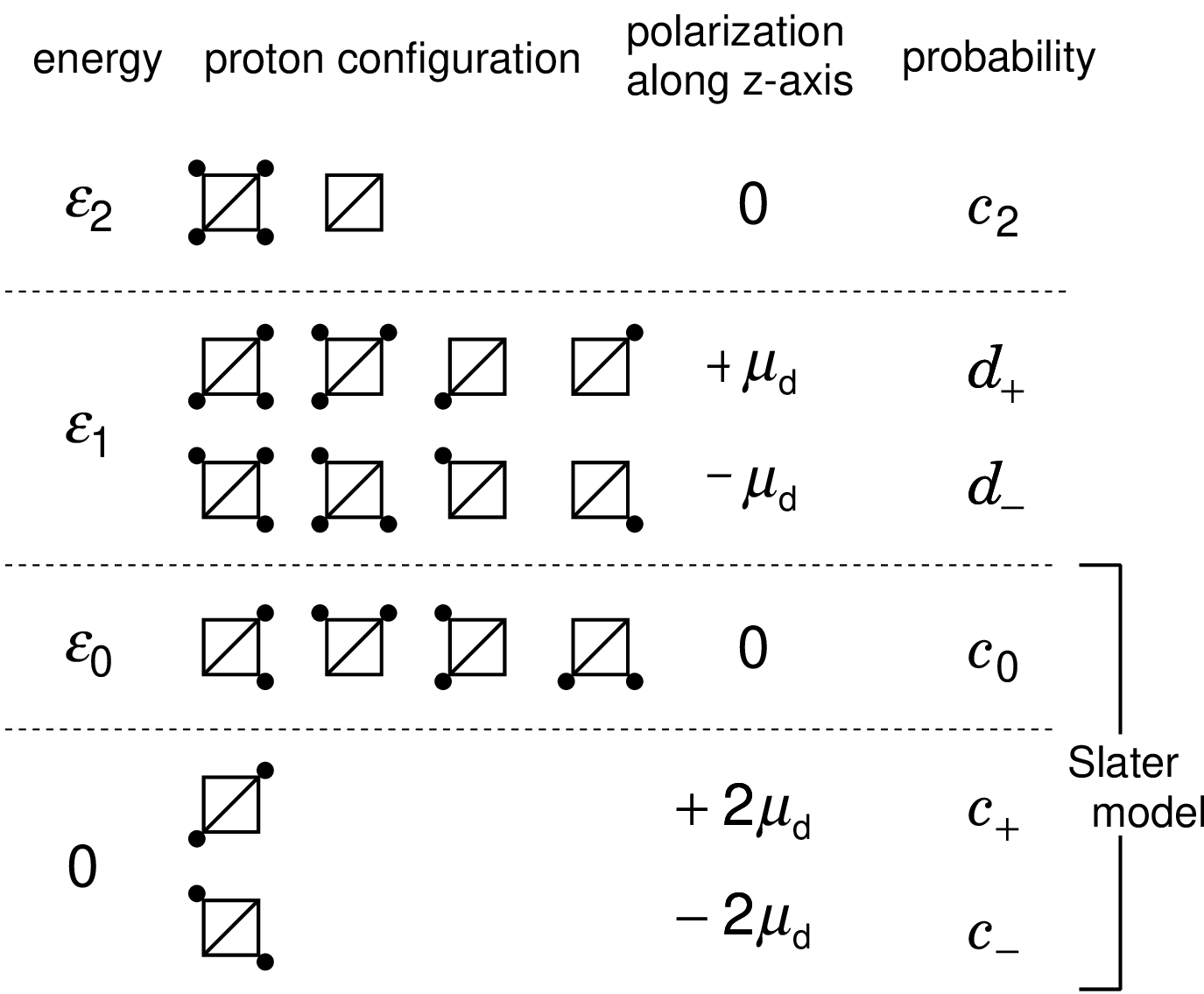,width=8cm}
\end{center}
\caption{Energy levels, magnitude of dipole moments of a K-PO$_4$
 complex and probability for proton configuration in Slater-Takagi
 model.} 
\label{fig:STmodel}
\end{figure}

There are $N$ PO$_4$ tetrahedra and $2N$ protons around PO$_4$
tetrahedra in the KDP-type crystal as shown in Fig.~\ref{fig:KH2PO4}.
The pseudo-spin Ising Hamiltonian $H$ for a configuration of $2N$
protons has a form~\cite{Wada} 
\begin{equation}
H=\sum_{\left<ijkl\right>}\left[H_0(\sigma_i,\sigma_j,\sigma_k,\sigma_l)-\frac{\mu_{\mathrm d}}{2}E(\sigma_i+\sigma_j+\sigma_k+\sigma_l)\right]
\label{Hamiltonian}
\end{equation}
with
\begin{eqnarray}
H_0(\sigma_i,\sigma_j,\sigma_k,\sigma_l)
 &=& - V_2(\sigma_i\sigma_j+\sigma_j\sigma_k+\sigma_k\sigma_l+\sigma_l\sigma_i)\nonumber\\[-1mm]
&&  - V_5(\sigma_i\sigma_k+\sigma_j\sigma_l)
   - V_4 \sigma_i\sigma_j\sigma_k\sigma_l
   + C \nonumber \\
\label{Hamiltonian0}
\end{eqnarray}
where the sum $\left<ijkl\right>$ runs over four protons $ijkl$ around each PO$_4$
tetrahedron in the crystal, $\sigma_i=\pm 1$ stands for the site of the
$i$-th proton in the double well potential along the O-O bond (hydrogen
bond) between two nearest neighbor PO$_4$'s, $\mu_{\mathrm d}$ is the magnitude of
an electric dipole moment associated with a complex K-H$_2$PO$_4$ and
$E$ is an external electric field. As is seen in Fig.1, we use a
convention that when the $i$-th proton is located on the  closer site to
an O atom at the top ( bottom ) of the PO$_4$ tetrahedron along the easy
z-axis, the $i$-th proton takes $\sigma_i=+1(-1)$. The energy parameters
$V_2,V_4,V_5$ and $C$ are related to those of the Slater-Takagi model
shown in Fig.~\ref{fig:STmodel} as
\begin{eqnarray}
&& V_2=\varepsilon_2/8\,,\quad
V_4=-\varepsilon_0/4 +\varepsilon_1/2 -\varepsilon_2/8\,, \nonumber\\[-3mm]
&& \\[-3mm]
&& V_5=\varepsilon_0/4 -\varepsilon_2/8\,,\quad
C=\varepsilon_0/4 +\varepsilon_1/2 +\varepsilon_2 \nonumber
\end{eqnarray} 
where the pseudo-spin Hamiltonian (\ref{Hamiltonian}) is a modified one
used by Tokunaga {\it et al.}~\cite{Tokunaga} by allowing all the
configurations of four protons around a PO$_4$ tetrahedron. Here the
ice-rule limit characteristic of the Slater model for KDP is realized
when $(\varepsilon_1-\varepsilon_0)/\varepsilon_0 \to \infty$ with $\varepsilon_2
>\varepsilon_1$. The ice rule is described by proton configurations in
which (A) only one proton exists on each hydrogen bond between two
nearest neighbor PO$_4$ tetrahedra and (B) all the PO$_4$ tetrahedra
have exactly two protons adjacent to them.

We now apply the path probability method(PPM) in the cactus approximation to 
the present system to find the kinetic equation for protons. The cactus
approximation equivalent to Slater's treatment~\cite{Slater} takes
account of the proton correlations around PO$_4$ as well as the site of
a proton in the double well potential on each hydrogen bond. However,
since the derivation of the kinetic equation by the PPM is a little
lengthy, though the final kinetic equation is relatively simple, here we
only mention the idea of the PPM~{\cite{KikuchiP, Mexico}}. In
equilibrium statistical mechanics, the realized state of a system in
thermal contact with a heat reservoir is the minimum state of its free
energy. When the system is not in equilibrium, we are interested in the
time evolution of the system. The PPM is a method for determining the
time evolution of the system. The idea of the PPM is to calculate a
transition probability of the ensemble of equivalent systems in a short
time interval ${\mathit\Delta} t$ from time $t$ to $t+{\mathit\Delta}
t$. This transition probability is called the path probability
function. Then, we assume an extremum principle that the path maximizing
this path probability function determines the time evolution of the
system.

Now, in the cactus approximation of the PPM, the homogeneous state of
the present system at time $t$ is described by five state variables
defined by  
\begin{eqnarray}
&&
m(t) = \left< \sigma_i \right>_t\,,\quad
s(t) = \left< \sigma_i\sigma_j\sigma_k \right>_t\,,\quad
q(t) = \left<\sigma_i\sigma_j \right>_t \,,
 \nonumber\\[2mm]
&&
q_{\mathrm D}(t)= \left<\sigma_i\sigma_k \right>_t \,,\quad
q_4(t) = \left<\sigma_i\sigma_j\sigma_k\sigma_l \right>_t
\end{eqnarray}   
where each state variable represents the correlation of protons $ijkl$
around a PO$_4$ cluster at time $t$(Fig.1) and $\left<\cdots \right>_t$ is an
thermal average at time $t$. After some manipulations of the PPM we
obtain a generating function from which a set of kinetic equations are
derived through differentiation of interaction parameters. The
generating function is given by 
\begin{eqnarray}
\lefteqn{G(\mib{L})=\theta\,\mathop{\mbox{\large Tr}}_ip_1(\sigma_i,t)e^{-2L_1\sigma_i}}\qquad && \nonumber\\[-1mm]
& &\times \left[\mathop{\mbox{\large Tr}}_{jkl}\frac{p_4(\sigma_i,\sigma_j,\sigma_k,\sigma_l,t)\,e^{{}-\frac{\beta}{2}
{\mathit\Delta}_{i}H_0(\sigma_i,\sigma_j,\sigma_k,\sigma_l)}}{p_1(\sigma_i,t)}\right]^2 \nonumber\\
\label{gen.fn}
\end{eqnarray}
\vspace{-5mm}
\begin{eqnarray*}
\lefteqn{{\mathit\Delta}_{i}H_0(\sigma_i,\sigma_j,\sigma_k,\sigma_l)}\qquad\qquad &&\nonumber\\
&=&
H_0(-\sigma_i,\sigma_j,\sigma_k,\sigma_l) - H_0(\sigma_i,\sigma_j,\sigma_k,\sigma_l)
\nonumber
\end{eqnarray*}
where $\mathop{\rm Tr}_i$ and $\mathop{\rm Tr}_{jkl}$ denote a trace
operation $\sum_{\sigma_i=\pm 1}$ and
$\sum_{\sigma_j,\sigma_k,\sigma_l=\pm 1}$, respectively,
$\beta=1/k_{\mathrm B}T$ is the inverse temperature, $\theta{}^{-1}$ is
a microscopic relaxation time of an isolated proton,
$L_1=\beta\mu_{\mathrm d} E/2$ and ${\mathit\Delta}_i$ defines an energy
increase under an inversion of only $\sigma_i$ variable into
$-\sigma_i$. Further, $p_1(\sigma_i,t)$ and
$p_4(\sigma_i,\sigma_j,\sigma_k,\sigma_l,t)$ are, respectively, the
probability  of finding the site $\sigma_i$ of a proton in the $i$-th
bond and the probability of finding the sites
$\sigma_i,\sigma_j,\sigma_k,\sigma_l$ of protons $i,j,k,l$ around a
PO$_4$ and are given in terms of above defined five state variables by 
\begin{eqnarray}
\lefteqn{p_1(\sigma_i,t)=\frac{1}{2}\bigl(1+m(t)\sigma_i\bigr)\,,}&& \nonumber\\
&& p_4(\sigma_i,\sigma_j,\sigma_k,\sigma_l,t) \nonumber\\
&& \, = \frac{1}{2^4}\Bigl(
1
 + m(t)(\sigma_i+\sigma_j+\sigma_k+\sigma_l) \nonumber\\[-1mm]
&&\hspace{1.5cm} + q(t)(\sigma_i\sigma_j+\sigma_j\sigma_k+\sigma_k\sigma_l+\sigma_l\sigma_i) \nonumber\\[-1mm]
&&\hspace{1.5cm} + q_{\mathrm D}(t)(\sigma_i\sigma_k+\sigma_j\sigma_l)\nonumber\\[-1mm]
&&\hspace{1.5cm} + s(t)(\sigma_i\sigma_j\sigma_k
        +\sigma_j\sigma_k\sigma_l
        +\sigma_k\sigma_l\sigma_i
        +\sigma_l\sigma_i\sigma_j)\nonumber\\[-1mm]
&&\hspace{1.5cm} + q_4(t)\sigma_i\sigma_j\sigma_k\sigma_l
\Bigr)\,.\nonumber\\
 \label{eqn:p4}
\end{eqnarray}
Then a set of kinetic equations are  given in a convenient form:~\cite{Kaburagi}
\begin{eqnarray}
\frac{dm_i(t)}{dt} = 4 \lim_{L_3\to 0} \frac{\partial G({\mib L})}{\partial L_i}
\quad\quad (i=1\sim 5)\label{kineticeq.}
\end{eqnarray}
Here, it should be noted that in order to write the above expression an
extra interaction term is virtually added to Hamiltonian (\ref{Hamiltonian0}) as 
\begin{eqnarray}
\lefteqn{H_0(\sigma_i,\sigma_j,\sigma_k,\sigma_l)}\hspace{1cm}&&\nonumber\\[-1mm]
&& - V_3(\sigma_i\sigma_j\sigma_k+\sigma_j\sigma_k\sigma_l+\sigma_k\sigma_l\sigma_i+\sigma_l\sigma_i\sigma_j)\nonumber\\
&& \hspace{-1cm}\longrightarrow H_0(\sigma_i,\sigma_j,\sigma_k,\sigma_l)\label{extra}
\nonumber\\   
\end{eqnarray}
and $V_3$ is, however, put to zero just after differentiation with respect to
$L_3=\beta V_3$ in eq.(\ref{kineticeq.}). We also redefine order
parameters as $m_1(t)=4m(t), m_2(t)=4q(t),m_3(t)=4s(t),m_4(t)=q_4(t)$
and $m_5(t)=2q_{\mathrm D}(t)$ and the corresponding fields as $L_1(t)=\beta\mu_b
E(t)/2, L_2=\beta V_2, L_3=\beta V_3, L_4 =\beta V_4$ and $L_5 =\beta
V_5$, respectively.

\section{Thermal equilibrium}
\begin{figure}[tbp]
 \hspace{-1.0cm}
 \epsfile{file=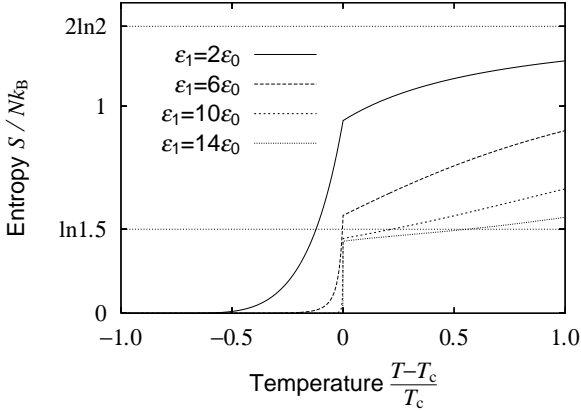,width=9cm}
\caption{Temperature dependence of the entropy for
  $\varepsilon_1=2\varepsilon_0,\,6\varepsilon_0,\,10\varepsilon_0,\,14\varepsilon_0$ and $\varepsilon_2=4\varepsilon_1-2\varepsilon_0$\,.}
\label{fig:entropy}
\end{figure}

In order to obtain the dynamical susceptibility $\chi{(\omega )}$ as the
linear response to the external field, equilibrium values of the order
parameters are required. Since the equilibrium state is more easily
obtained from the CVM than from the stationary solution of the kinetic
equation $(\ref{kineticeq.})$, we apply the cactus approximation of the
CVM to the present system.~\cite{Wada,Ishibashi} The variational
free energy $F$ is obtained  by 
\begin{eqnarray}
F= U -TS 
\end{eqnarray}
where the internal energy $U$ is given by
\begin{equation}
U/N = -4 V_2 q - 2 V_5 q_{\mathrm D} - V_4 q_4 - 2 \mu_{\mathrm d} m E
\end{equation} 
and the entropy $S$ is given by 
\begin{eqnarray}
\lefteqn{
S/Nk_{\mathrm{B}}
 = 2 \mathop{\mbox{Tr}}_{i} \,\Bigl[p_1(\sigma_i)\ln{p_1(\sigma_i)}\Bigr]
}\hspace{1.5cm}&& \nonumber\\
 && - \mathop{\mbox{Tr}}_{ijkl} \Bigl[p_4(\sigma_i,\sigma_j,\sigma_k,\sigma_k)\ln{p_4(\sigma_i,\sigma_j,\sigma_k,\sigma_l)}\Bigr]. \nonumber\\
\label{entropy}
\end{eqnarray}
The minimum condition of the variatinal free energy with respect to $m,
s,q, q_{\mathrm D},q_4$ yields equilibrium relations with a definition
$h_{\mathrm e}=\exp{(\beta\mu_{\mathrm d} E)}$ 
\begin{equation}
\begin{array}{c}
\displaystyle
\frac{c_+}{c_-} = \left(\frac{1+m}{1-m}\right)^2 h_{\mathrm e}{}^4 \,,\quad
\frac{d_+}{d_-} = \left(\frac{1+m}{1-m}\right) h_{\mathrm e}{}^2 \,,\\[3mm]
\displaystyle
\frac{c_2}{c_0} = \frac{\eta_2}{\eta_0} \,,\quad
\frac{c_+ c_-}{c_0{}^2} = \frac{1}{\eta_0{}^2} \,,\quad
\frac{d_+ d_-}{c_0{}^2} = \left(\frac{\eta_1}{\eta_0}\right)^2
\end{array}
\label{ratio}
\end{equation}
where $c_{+},\, c_{-},\, d_{+},\, d_{-},\, c_2$ and $c_0$ shown in
fig.\ref{fig:STmodel} are defined from
$p_4(\sigma_i,\sigma_j,\sigma_k,\sigma_l)$ by 
\begin{eqnarray}
\begin{array}{l}
c_+ = \displaystyle\frac{1}{2^4}(1+4m+4q+2q_{\mathrm D}+4s+q_4)\,, \\[3mm]
c_- = \displaystyle\frac{1}{2^4}(1-4m+4q+2q_{\mathrm D}-4s+q_4)\,, \\[3mm]
d_+ = \displaystyle\frac{1}{2^4}(1+2m-2s-q_4)\,, \\[3mm]
d_- = \displaystyle\frac{1}{2^4}(1-2m+2s-q_4)\,, \\[3mm]
c_2 = \displaystyle\frac{1}{2^4}(1-4q+2q_{\mathrm D}+q_4)\,, \\[3mm]
c_0 = \displaystyle\frac{1}{2^4}(1-2q_{\mathrm D}+q_4)
\end{array}
\label{definition}
\end{eqnarray}
with a normalization condition
\begin{eqnarray}
c_{+}+c_{-}+4c_0+4(d_{+}+d_{-})+2c_2 =1 \label{normalization}.
\end{eqnarray}
From eq.(\ref{ratio}) and eq.(\ref{normalization}) it is  easy to solve
$c_{+},c_{-},d_{+},d_{-},c_2,c_0$ in terms of the polarization $m$ and
field variable $h_{\mathrm e}$: 
\begin{full}
\begin{eqnarray}
&&
c_0=\frac{\eta_0}{4\eta_0+2\eta_2+(h_{\mathrm e}{}^2\frac{1+m}{1-m}+h_{\mathrm e}{}^{-2}\frac{1-m}{1+m})+4\eta_1(h_{\mathrm e}\sqrt{\frac{1+m}{1-m}}+h_{\mathrm e}{}^{-1}\sqrt{\frac{1-m}{1+m}})}\,,\nonumber\\
&&
c_{+}=h_{\mathrm e}{}^2\frac{1+m}{1-m}\frac{c_0}{\eta_0}\,,\quad
c_{-}=h_{\mathrm e}{}^{-2}\frac{1-m}{1+m}\frac{c_0}{\eta_0}\,,\quad
c_2=\frac{\eta_2}{\eta_0}c_0 \,, \label{mkai}\\
&&
d_{+}=h_{\mathrm e}\sqrt{\frac{1+m}{1-m}}\frac{\eta_1c_0}{\eta_0}\,,\quad
d_{-}=h_{\mathrm e}{}^{-1}\sqrt{\frac{1-m}{1+m}}\frac{\eta_1c_0}{\eta_0}
\nonumber
\end{eqnarray}
\end{full}
The spontaneous polarization $m_0$~\cite{Ishibashi} is determined by the
relation 
\begin{eqnarray}
m
 = p_1(+1)-p_1(-1)
 = c_+ - c_- + 2 (d_+ - d_- ) \nonumber\\[-3mm]
\end{eqnarray} 
as 
\begin{equation}
m_0 = \left\{
\begin{array}{cl}
\displaystyle\sqrt{1 - \frac{4\eta_1^2}{(1-2\eta_0-\eta_2)^2}} & \quad \mbox{for } T<T_{\mathrm C}  \\[7mm]
0& \quad  \mbox{for } T>T_{\mathrm C}\\[2mm]
\end{array}\label{order} \right.
\end{equation}
and the electric susceptibility is given by~\cite{Wada} 
\begin{equation}
\chi_{\mathrm stat} =\frac{\mu_{\mathrm d}^2}{k_{\mathrm B}T}\frac{2(1+\eta_1(1-2m_0^2)/\sqrt{1-m_0^2})}
{-1+2\eta_0+\eta_2+2\eta_1/\sqrt{(1-m_0^2)^3}}
\end{equation}
The other equilibrium order parameters in eq.(\ref{mkai}) without
electric field are found from eq.(\ref{order}). Further, utilizing
eq.(\ref{entropy}) the entropy versus temperature are shown in
Fig.\ref{fig:entropy}, in order to compare our result with experimental
data of Matsuo {\it et al}~\cite{Matsuo}. In the following figures energy parameters such as
$\varepsilon_1$ are measured in the units of Boltzmann factor $k_{\mathrm
B}$.

\section{Dynamical response in paraelectric phase}
In the paraelectric phase we can carry out all the calculations analytically. 
In thermodynamic equilibrium of the paraelectric phase, under the inversion of external field $E$, the order parameters $m$ and $s$ are changed into $-m$ and $-s$, respectively, while $q$, $q_{\mathrm D}$ and $q_4$ are  invariant. Thus $m(t)$ and $s(t)$ are long range order parameters responding linearly to external field $E(t)$, while $q(t), q_{\mathrm D}(t)$ and $q_4(t)$ are short range order parameters responding quadratically to the field. Then, in order to obtain the linear dynamical 
susceptibility of the present system above the transition temperature,
the short range order parameters $q(t), q_{\mathrm D}(t), q_4(t)$ can be
replaced by the values at thermal equilibrium in paraelectric phase
without electric field. A set of kinetic equations $(\ref{kineticeq.})$ with five equations  is reduced  to two closed equations for long range order parameters $m(t)$ and $s(t)$ up to a linear order to the external field by     
\begin{full}
\begin{eqnarray}
\frac{dm(t)}{dt}
 &=& 8\theta\lambda^2\left[
L_1
 + \left(1-\frac{2\eta_1+2\sqrt{\eta_0}+\sqrt{\eta_2}-1}{4\sqrt{\eta_1}\lambda}\right) m(t)
 - \frac{2\eta_1-2\sqrt{\eta_0}-\sqrt{\eta_2}+1}{4\sqrt{\eta_1}\lambda}s(t)
\right]\,,\nonumber\\[2mm]
\frac{ds(t)}{dt}
 &=& 8\theta\lambda\mu \left[
 L_1 + \left(1-\frac{2\eta_1+2\sqrt{\eta_0}+\sqrt{\eta_2}-1}{8\sqrt{\eta_1}\lambda}+\frac{-6\eta_1+2\sqrt{\eta_0}+\sqrt{\eta_2}+3}{8\sqrt{\eta_1}\mu}\right)m(t)\right.\\
&& \hspace{3cm}  - \left.\left(\frac{2\eta_1-2\sqrt{\eta_0}-\sqrt{\eta_2}+1}{8\sqrt{\eta_1}\lambda}
+\frac{6\eta_1+2\sqrt{\eta_0}+\sqrt{\eta_2}+3}{8\sqrt{\eta_1}\mu}\right) s(t) \right]
\nonumber
\end{eqnarray}
\end{full}
with $\eta_0=e{}^{-\beta\varepsilon_0},\,\eta_1=e{}^{-\beta\varepsilon_1},\,
\eta_2=e{}^{-\beta\varepsilon_2}$ and 
\begin{eqnarray}
\lambda
 &=& \frac{\sqrt{\eta_1}(1+2\sqrt{\eta_0}+\sqrt{\eta_2})}{1+2\eta_0+4\eta_1+\eta_2} \\
\mu
 &=& \frac{\sqrt{\eta_1}(3-2\sqrt{\eta_0}-\sqrt{\eta_2})}{1+2\eta_0+4\eta_1+\eta_2}.
\end{eqnarray} 
Now we assume $\mu_{\mathrm d} m(t)=\chi (\omega )E\exp{i\omega t}$ and $\mu_{\mathrm d} s(t)=\chi_{s}(\omega )E\exp{i\omega t}$. Substituting these relations into this set of kinetic equations, we finally obtain the dynamical susceptibility $\chi (\omega )$ by
\begin{full}
\begin{eqnarray}
\chi (\omega ) =\frac{\mu_{\mathrm d}^2}{k_{\mathrm B}T}\frac{\theta\lambda}{A_{+}+A_{-}}
\left(
\frac{D-8A_{-}\lambda}{i\omega+\theta A_{-}\lambda}
-\frac{D-8A_{+}\lambda}{i\omega+\theta A_{+}\lambda}
\right)
\label{chiomega}
\end{eqnarray}
\end{full}
where $A_{\pm},D$ are given as
\begin{full}
\begin{eqnarray}
&&
A_{\pm}=B \pm\sqrt{B^2-C}\,,\quad
B=2 \left(
\frac{1+4\eta_0+4\eta_1+\eta_2+4\eta_1\sqrt{\eta_0}+4\sqrt{\eta_0\eta_2}+2\eta_1\sqrt{\eta_2}}{\sqrt{\eta_1}(1+2\sqrt{\eta_0}+\sqrt{\eta_2})}-2\lambda
\right)\,,\nonumber\\[-3mm]
\\[-3mm]
&&
C=\frac{32(2\sqrt{\eta_0}+\sqrt{\eta_2})(-1+2\eta_0+2\eta_1+\eta_2)}{1+2\eta_0+4\eta_1+\eta_2}\,,\quad
D=\frac{64(1+\eta_1)(2\sqrt{\eta_0}+\sqrt{\eta_2})}{1+2\eta_0+4\eta_1+\eta_2}\,.\nonumber
\end{eqnarray}
\end{full} 
Especially the static susceptibility $\chi_{\mathrm stat}$ is obtained by putting $\omega =0$ in eq.(\ref{chiomega}) as 
\begin{eqnarray}
\chi_{\mathrm stat}=\chi{(\omega =0)} =\frac{\mu_{\mathrm d}^2}{k_{\mathrm B}T}
\frac{2(1+\eta_1)}{-1+2\eta_0+2\eta_1+\eta_2}
\end{eqnarray}
The transition temperature $T_{\mathrm c}$ to the ferroelectric phase is determined as a 
divergent point of the static susceptibility as 
\begin{eqnarray}
2\, e{}^{-\varepsilon_0/k_{\mathrm B}T_{\mathrm c}}
 + 2\, e{}^{-\varepsilon_1/k_{\mathrm B}T_{\mathrm c}}
 + e{}^{-\varepsilon_2/k_{\mathrm B}T_{\mathrm c}}
 = 1 
\end{eqnarray}
This expression  is the one obtained by Ishibashi~\cite{Ishibashi}. 
When the transition temperature is approached, the first term of
eq.(\ref{chiomega}) shows a critical slowing down and contributes mainly
to the dynamical susceptibility.

\section{Dynamical response in ferroelectric phase}
\begin{figure}[tbp]
\hspace{-1.0cm}\epsfile{file=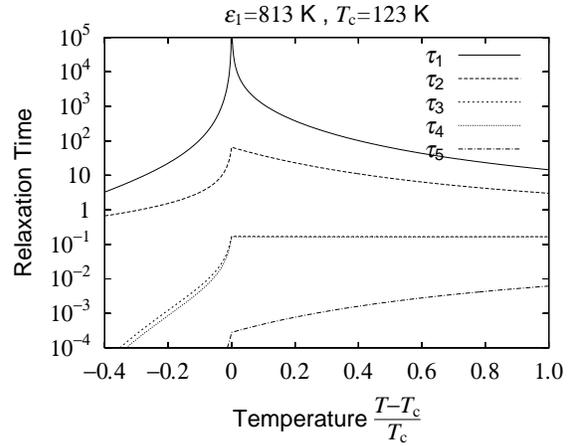,width=9cm}
\caption{Relaxation times versus temperature for
 $\varepsilon_1=813\,\mbox{K}$. For $T>T_{\mathrm c}$,
 only $\tau_1$ and $\tau_2$ take part in $\chi(\omega)$.} 
\label{fig:tau}
\end{figure}

\begin{figure}
\epsfile{file=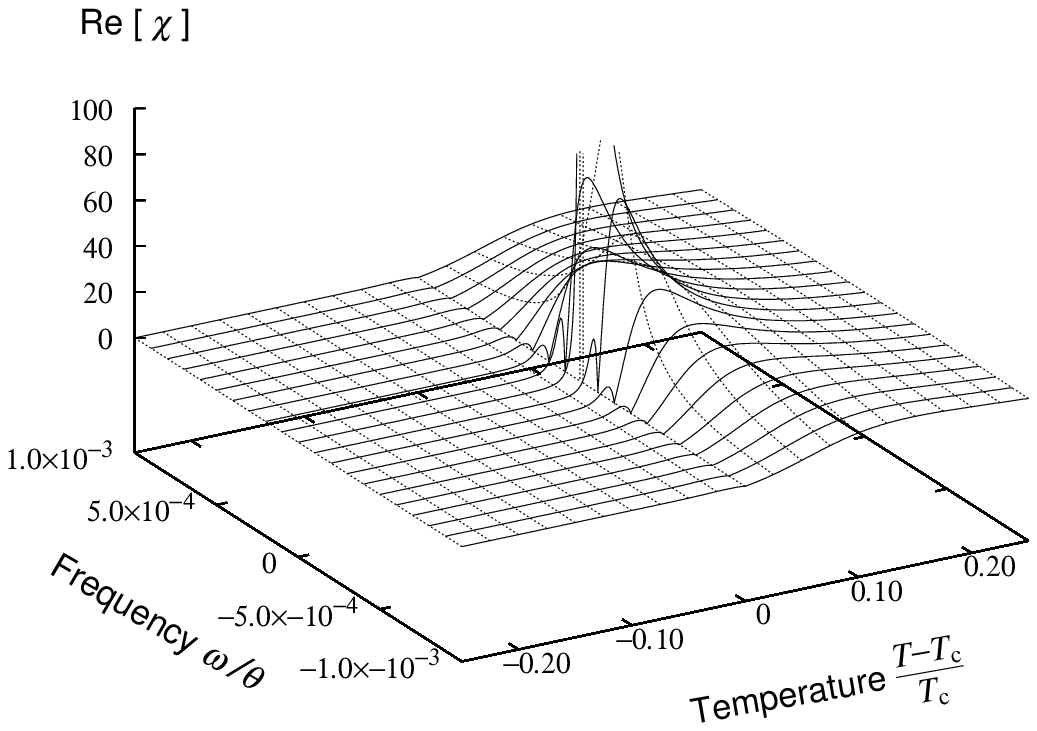,width=9cm}

\epsfile{file=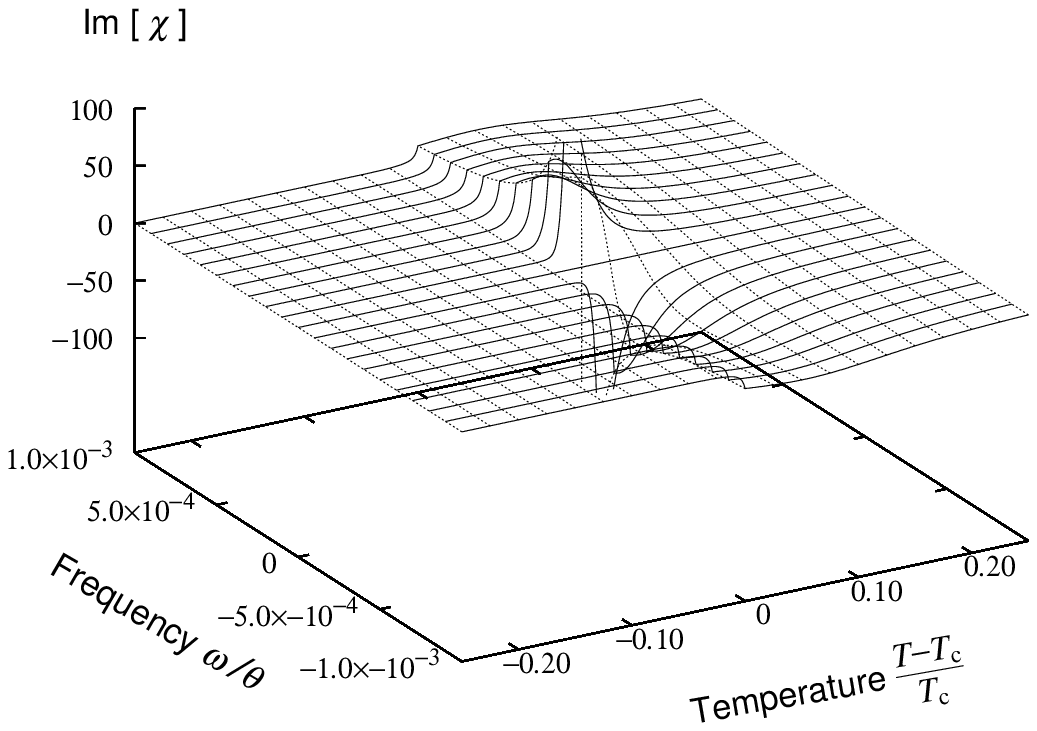,width=9cm}
\caption{$\omega$ and $(T-T_{\mathrm c})/T_{\mathrm c}$ dependence of
 $\chi'(\omega)$ and $\chi''(\omega)$ for $\varepsilon_0=85.6 \mbox{K},\,\varepsilon_1=813 \mbox{K},\,\varepsilon_2=4\varepsilon_1-2\varepsilon_0$. }
\label{fig:chi-omega-T}
\end{figure}

In the ferroelectric phase a finite spontaneous polarization $m_0$ occurs 
as given in eq.(\ref{order}). 
When the external electric field $E(t)$ is applied, the short range order parameters $q(t), q_{\mathrm D}(t),q_4(t)$ have also components proportional to $E(t)$ indirectly through the spontaneous polarization $m_0$. Thus we assume that
\begin{eqnarray}
m_1(t)&=&4m(t)=4m_0+\chi_1{(\omega )}E\exp{i\omega t}\nonumber\\
m_2(t)&=& 4q(t)=4q_0+\chi_2{(\omega )}E\exp{i\omega t}\nonumber\\  
m_3(t)&=&4s(t)=4s_0+\chi_3{(\omega )}E\exp{i\omega t} \label{displacement}\\
m_4(t)&=&q_4(t)=q_4^0+\chi_4{(\omega )}E\exp{i\omega t}\nonumber\\
m_5(t)&=&2q_{\mathrm D}(t)=2q_{\mathrm D}^0+\chi_5{(\omega )}E\exp{i\omega t}\nonumber
\end{eqnarray}
where the required dynamical susceptibility is $\chi{(\omega
)}=\mu_{\mathrm d} \chi_1(\omega)/4$ and $m_0,s_0,q_0, q_{\mathrm D}^0$
and $q_4^0 $ are equilibrium order parameters in the absence of external
electric field. Then, substituting eq.(\ref{displacement}) into
eq.$(\ref{kineticeq.})$, we finally obtain a set of algebraic equations
for five $\chi_i(\omega )\quad (i=1\sim 5)$ which is read in a matrix
form as 
\begin{equation}
\left(
\frac{i\omega}{\theta}I+M \right) \mib{\chi}(\omega) \label{matrix}
 = \mib{b}
\end{equation} 
Since the explicit forms of a matrix $M$ and a column vector $\mib{b}$
are  complicated and lengthy compared with the paraelectric phase, they
are given in Appendix. Referring to Appendix, the elements of the matrix
$M$ and the column vector $\mib{b}$ can be expressed in terms of only
$\eta_0,\eta_1$ and $\eta_2$ without unknown state variables since order
parameters at equilibrium are obtained analytically. Then the
algebraic equation can be easily calculated numerically for fixed
$\omega$ using the Gaussian elimination method for linear algebraic
equation. With the relaxation time $\tau_i\quad (i=1\sim 5)$, the
dynamical susceptibility per proton $\chi (\omega )$ can be written for
$T<T_{\mathrm c}$ as
\begin{equation}
\chi{(\omega)}
 = \frac{\mu_{\mathrm d}}{4} \chi_1(\omega )
 = \sum_{i=1}^5\frac{\chi_i}{1+i\omega \tau_i}
\end{equation} 
where the relaxation times $\tau_i$ are obtained by a diagonalization of 
$M$ in terms of a matrix $U$ as 
\begin{eqnarray}
&&(UMU^{-1})_{ij} = \frac{1}{\theta \tau_i}\,\delta_{ij}\\[1mm]
&& \qquad\qquad (\delta_{ij} : \mbox{Kronecker's delta}) \nonumber
\end{eqnarray}
and the intensity coefficients $\chi_i$ are given by
\[
\chi_i = \frac{1}{4} \theta \mu_{\mathrm d} \tau_i \sum_{j=1}^5
(U^{-1})_{1i}U_{ij} b_j \,.
\]
Especially for $T>T_{\mathrm c}$, $\chi_3, \chi_4$ and $\chi_5 $
representing the intensity from each relaxation mode reduce to zero and
in consistency to eq.(\ref{chiomega}) we obtain  
\begin{eqnarray}
\chi{(\omega )} =\sum_{i=1}^2\frac{\chi_i}{1+i\omega\tau_i}\,.
\end{eqnarray}
Though there appear five relaxation times in the ferroelectric phase,
only  one of them shows the critical slowing down and contributes mainly
to the dynamical susceptibility when the transition temperature is
approached (Fig.\,\ref{fig:tau}).
The real $\chi^{'}(\omega )$ and imaginary
$\chi^{''}(\omega )$ part of the dynamical susceptibility are defined as
$\chi{(\omega )}\equiv \chi^{'}(\omega)-i\chi^{''}(\omega)$. The results
for  $\chi^{'}(\omega )$ and $\chi^{''}(\omega )$ versus $\omega$ and
temperature $(T-T_{\mathrm c})/T_{\mathrm c}$ in the para- and
ferro-electric phases are shown in Fig.\ref{fig:chi-omega-T}.

\section{Results and discussions}
The result shows that, though there are, respectively, two and five
relaxation times in para- and ferro-electric phase according to a set of
independent kinetic equations,  only one of the relaxation times shows a
critical slowing down when the temperature $T$ approaches the transition
temperature $T_{\mathrm c}$ from above and below the transition
temperature and makes a main contribution to the dynamical
susceptibility.
 
\begin{figure}[btp]
\hspace{1mm}\epsfile{file=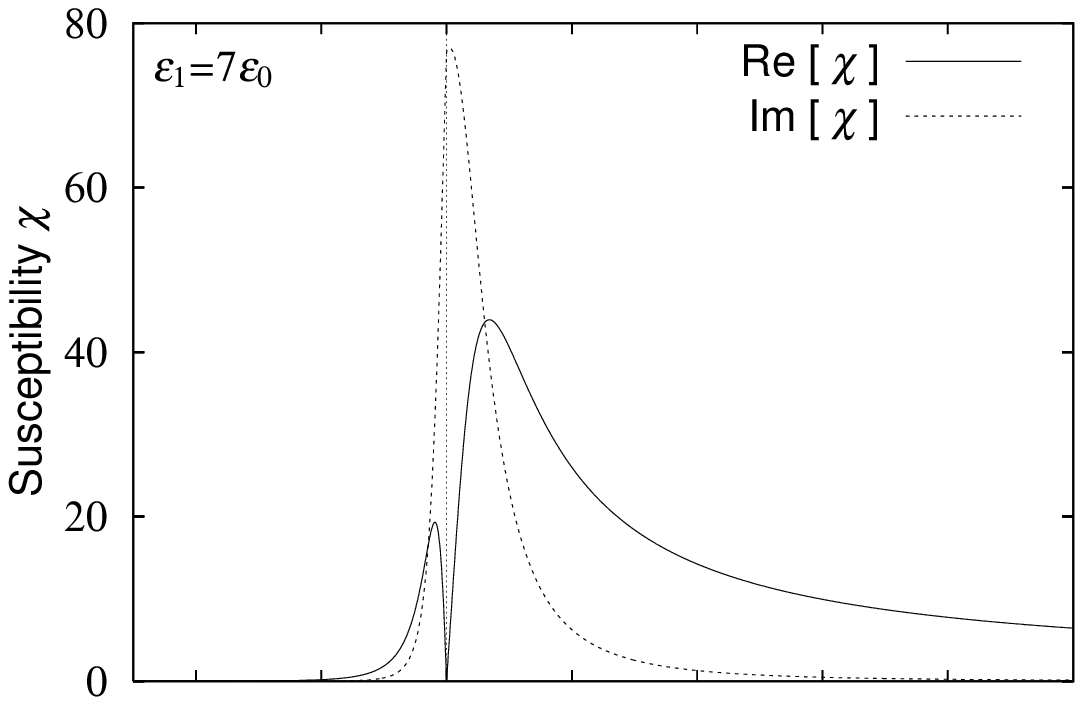,width=8cm}
\vspace{-1cm}

\epsfile{file=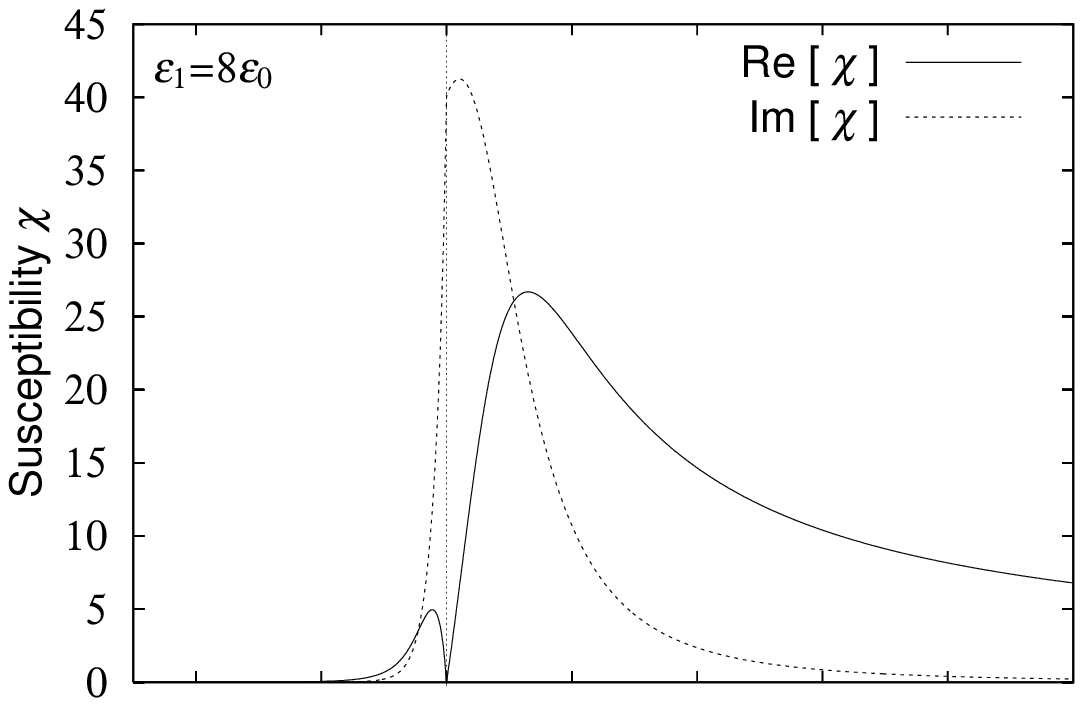,width=8cm}
\vspace{-1cm}

\epsfile{file=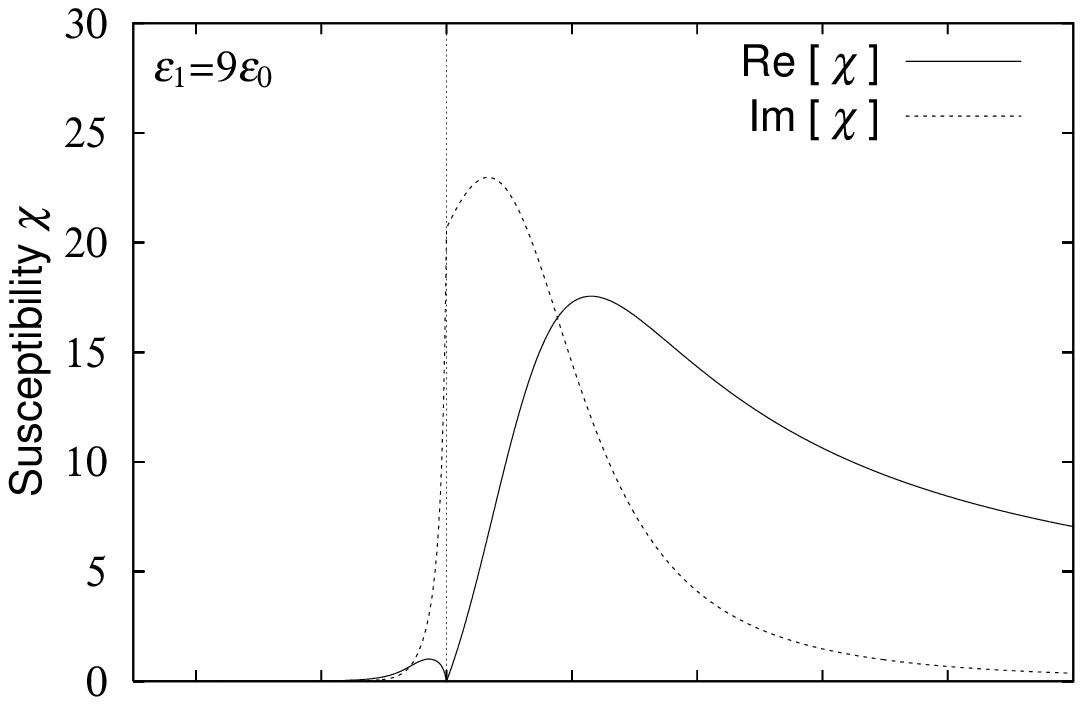,width=8cm}
\vspace{-1cm}

\epsfile{file=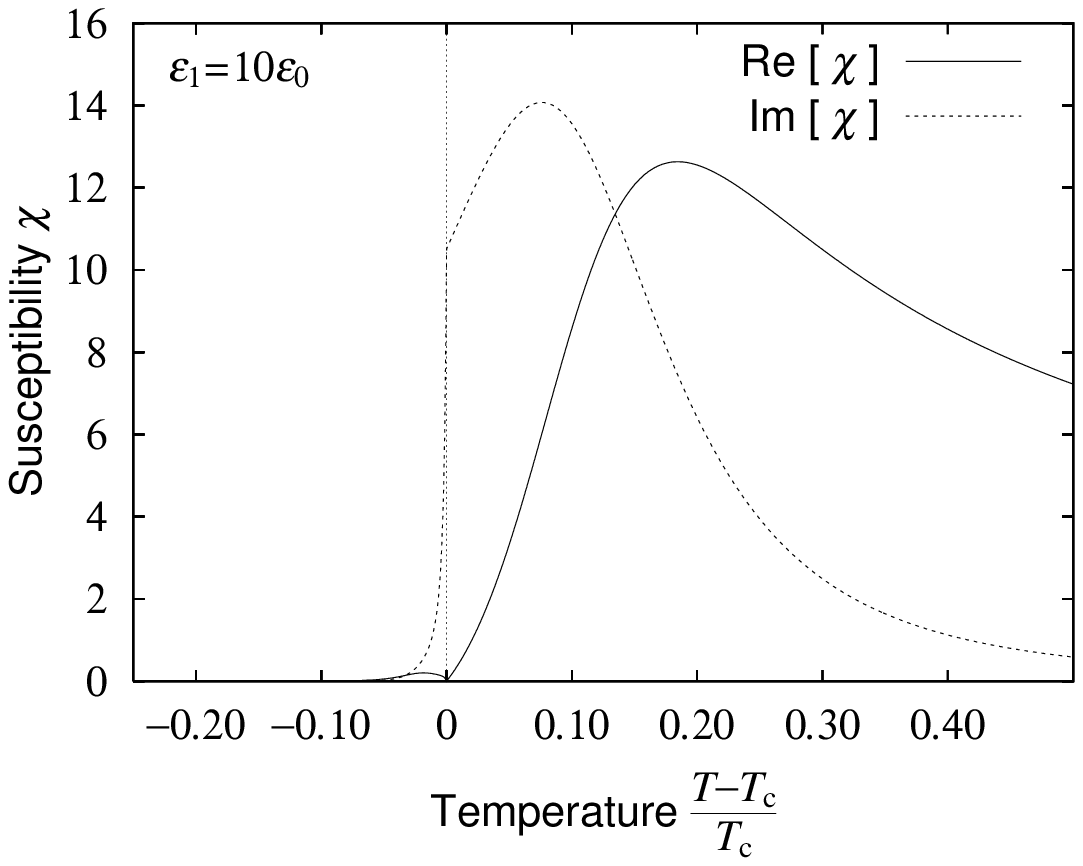,width=8cm}
\caption{Temperature dependence of the susceptibility $\chi(\omega)$ for
 various $\varepsilon_1$ under the constant $\omega$. The solid and
 dotted curve denotes, respectively, the real $\chi'(\omega)$ and the
 imaginary part $\chi''(\omega)$}
\label{fig:chi-T-e1}
\end{figure}

In order to compare the experimental data with our results we present
the temperature dependence of the real part $\chi'{(\omega )}$ and
imaginary part $\chi''(\omega)$ for constant $\omega$
for various values $\varepsilon_1$ (Fig.\ref{fig:chi-T-e1}).
The hilly behaviors appear not only above the transition temperature
 $T_{\mathrm c}$ but also below $T_{\mathrm c}$ for the finite
 $\varepsilon_1$ whereas in the ice-rule limit $\chi{(\omega )}$
 vanishes completely below $T_{\mathrm c}$. The dip of $\chi^{'}{(\omega
 )}$ at $T=T_{\mathrm c}$ is caused by the vanishing of the numerator
 due to the contribution from the relaxation mode showing a critical
 slowing down. The experimental data~\cite{Kozlov} show the hilly
 behavior below the transition temperature and the dip structure at the
 transition  temperature. In our calculation the contribution from the
 relaxation mode showing a critical slowing down overwhelms
 contributions from other modes and the dip goes to almost zero contrary
 to the experimental data. Recently, Matsuo {\it et al.}~\cite{Matsuo}
 re-examined experiments of the heat capacity and  estimated the
 transition entropy ${\mathit \Delta} S$ due to proton ordering from the
 experimental data. They discussed the discrepancy from the Slater
 theory and estimated the contribution from the excitation level of the
 Slater-Takagi model.We presented  the entropy curve versus temperature
 from our calculation for various parameters in
 fig.\ref{fig:entropy}. These results reveal that the ice-rule in the
 Slater model is not completely satisfied in the KDP crystal.

The dynamical susceptibility has been calculated not only above the
transition temperature but also below it in the cactus approximation of
the Slater-Takagi model utilizing an analytical solution  for the
spontaneous polarization. The results based on the Slater-Takagi model
are in good agreement with the experiments on dynamical susceptibility
and excess entropy.

\section*{Acknowledgements}
We would like to thank Prof. Tokunaga for an introduction to KDP-type
 ferroelectrics and many intensive discussions. We are also grateful to
 members of the statistical physics group for various discussions. 

\appendix
\section{Derivation of eq. (\ref{matrix})}
The generating function ($\ref{gen.fn}$) is conveniently written as 
\begin{eqnarray}
G(\mib{L})=\theta\mathop{\mbox{\large Tr}}_i 
\frac{2}{1+m(t)\sigma_i}
\left(
\mathop{\mbox{\large Tr}}_{jkl}
p_4(\{\sigma\}_{ijkl}) e^{\mib{L}\cdot{\mathit \Delta}_i\mib{\sigma}/2}
\right)^2
\nonumber\\
\end{eqnarray}
where $\mib{L}$ is the energy parameter vector defined as
\begin{equation}
\mib{L}
 = \beta 
\bigl(
C \quad \mu_{\mathrm d} E/2 \quad V_2 \quad V_3 \quad V_4 \quad V_5
\bigr)
\end{equation}
and the Ising spin vector $\mib{\sigma}$ is defined as
\begin{equation}
\mib{\sigma}
 =  \left(\begin{array}{l}
1\\
\sigma_i+\sigma_j + \sigma_k+\sigma_l \\
\sigma_i\sigma_j + \sigma_j\sigma_k + \sigma_k\sigma_l + \sigma_l\sigma_i \\
\sigma_i\sigma_j\sigma_k + \sigma_j\sigma_k\sigma_l + \sigma_k\sigma_l\sigma_i + \sigma_l\sigma_i\sigma_j \\
\sigma_i\sigma_j\sigma_k\sigma_l \\
\sigma_i\sigma_k + \sigma_j\sigma_l               
\end{array}\right)\,.
\end{equation}
The thermal average of the Ising spin vector is the order parameter
vector defined by
\begin{equation}
\mib{m}(t)
 = 
\bigl(
1 \quad 4 m(t) \quad 4 q(t) \quad 4 s(t) \quad q_4(t) \quad 2 q_{\mathrm D}(t)
\bigr)\,.
\end{equation}
The probability
$p_4(\{\sigma\}_{ijkl})(\equiv p_4(\sigma_i,\,\sigma_j,\,\sigma_k,\,\sigma_l))$
can also be written by using $\mib{\sigma}$ and $\mib{m}(t)$ as
eq.(\ref{eqn:p4}). Then a set of kinetic equation $(\ref{kineticeq.})$
is written in a vector form by 
\begin{eqnarray}
\frac{d{\mathit \Delta}\mib{m}(t)}{dt} =4 \frac{\partial G(\mib{L})}{\partial\mib{L}}
\end{eqnarray}
and is further rewritten to a linear order of the external electric field $E$ as 
\begin{full}
\begin{eqnarray}
\frac{d{\mathit \Delta} \mib{m}(t)}{d(\theta t)}
 &=& -4\mu_{\mathrm d} E \beta \mathop{\mbox{\large Tr}}_i \frac{\sigma_i}{p_1^{\mathrm e}(\sigma_i)} \bigl(\mathop{\mbox{\large Tr}}_{jkl} p_4^{\mathrm e}(\{\sigma\}_{ijkl}) h(\{\sigma\}_{ijkl})\bigr) \bigl(\mathop{\mbox{\large Tr}}_{jkl} p_4^{\mathrm e}(\{\sigma\}_{ijkl})\mib{h}'(\{\sigma\}_{ijkl})\bigr) \nonumber\\
&& - 2{\mathit \Delta} m(t) \mathop{\mbox{\large Tr}}_i \frac{\sigma_i}{\left(p_1^{\mathrm e}(\sigma_i)\right)^2} \bigl(\mathop{\mbox{\large Tr}}_{jkl} p_4^{\mathrm e}(\{\sigma\}_{ijkl}) h(\{\sigma\}_{ijkl})\bigr) \bigl(\mathop{\mbox{\large Tr}}_{jkl} p_4^{\mathrm e} (\{\sigma\}_{ijkl}) \mib{h}'(\{\sigma\}_{ijkl})\bigr) \nonumber \\
&& + 4 \mathop{\mbox{\large Tr}}_i \frac{1}{p_1^{\mathrm e}(\sigma_i)} \bigl(\mathop{\mbox{\large Tr}}_{jkl} p_4^{\mathrm e}(\{\sigma\}_{ijkl})\mib{h}'(\{\sigma\}_{ijkl})\bigr) \bigl(\mathop{\mbox{\large Tr}}_{jkl} {\mathit\Delta} p_4^{\mathrm e}(\{\sigma\}_{ijkl}) h(\{\sigma\}_{ijkl})\bigr) \nonumber\\
&& + 4 \mathop{\mbox{\large Tr}}_i \frac{1}{p_1^{\mathrm e}(\sigma_i)} \bigl(\mathop{\mbox{\large Tr}}_{jkl} p_4^{\mathrm e}(\{\sigma\}_{ijkl}) h(\{\sigma\}_{ijkl})\bigr) \bigl(\mathop{\mbox{\large Tr}}_{jkl} {\mathit\Delta} p_4^{\mathrm e}(\{\sigma\}_{ijkl}) \mib{h}'(\{\sigma\}_{ijkl})\bigr) \label{linear}
\end{eqnarray}
\end{full} 
where $p_1^{\mathrm e}(\sigma)=(1+m_0\sigma )/2$ and $p_4^{\mathrm
e}(\{\sigma\}_{ijkl})=p_4^{\mathrm
e}(\sigma_i,\sigma_j,\sigma_k,\sigma_l)$ are thermal equilibrium
probabilities without external field $E=0$ and ${\mathit
\Delta}\mib{m}(t)$ is a linear difference from equilibrium value of
order parameter vector $\mib{m}(t)$. The variable $h$ and the vector
$\mib{h}'$ are further defined as
\begin{eqnarray}
&&
 h(\{\sigma\}_{ijkl})
 = h(\sigma_i,\sigma_j,\sigma_k,\sigma_l)
 = e^{\mib{L}_0\cdot{\mathit \Delta}_i\mib{\sigma}/2}\,, \nonumber\\
&&
 \mib{h}'(\{\sigma\}_{ijkl})
 = \mib{h}'(\sigma_i,\sigma_j,\sigma_k,\sigma_l)
 = e^{\mib{L}_0\cdot{\mathit \Delta}_i\mib{\sigma}/2}{\mathit \Delta}_i\mib{\sigma}\nonumber
\end{eqnarray}   
where $\mib{L}_0$ is an interaction parameter vector $\mib{L}$ with $E=0$.
In order to obtain the dynamical susceptibility $\chi{(\omega )}$,
${\mathit \Delta} \mib{m}(t) =\mib{\chi}{(\omega )}E\exp{i\omega t}$ is
assumed and then eq.(\ref{linear}) can be easily rewritten into a form
$(\ref{matrix})$ in the main text:
\begin{equation}
\left(
\frac{i\omega}{\theta}I+M \right) \mib{\chi}(\omega)
 = \mib{b}\,.
\end{equation}

\end{document}